# Effect of electric current on optical orientation of electrons in AlGaAs/GaAs heterostructure

O. S. Ken[1], E. A. Zhukov[1,2], I. A. Akimov[1,2], V. L. Korenev[1], N. E. Kopteva[2,3], I. V. Kalitukha[1], V. F. Sapega[1], A. D. Wieck[4], A. Ludwig[4], R. Schott[4], Yu. G. Kusrayev[1], D. R. Yakovlev[1,2], and M. Bayer[1,2]

[1]*Ioffe Institute, Russian Academy of Sciences, 194021 St. Petersburg, Russia*
[2]*Experimentelle Physik 2, Technische Universität Dortmund, D-44227 Dortmund, Germany*
[3]*Spin Optics Laboratory, St. Petersburg State University, 198504 St. Petersburg, Russia*
[4]*Fakultät für Physik und Astronomie, Ruhr-Universität Bochum, 44801 Bochum, Germany*

**Abstract**

The effect of a lateral electric current on the photoluminescence H-band of an AlGaAs/GaAs heterostructure is investigated. The photoluminescence intensity and optical orientation of electrons contributing to the H-band are studied by means of continuous wave and time-resolved photoluminescence spectroscopy and time-resolved Kerr rotation. It is shown that the H-band is due to recombination of the heavy holes localized at the heterointerface with photoexcited electrons attracted to the heterointerface from the GaAs layer. Two lines with significantly different decay times constitute the H-band: a short-lived high-energy one and a long-lived low-energy one. The high-energy line originates from recombination of electrons freely moving along the structure plane, while the low-energy one is due to recombination of donor-bound electrons near the interface. Application of the lateral electric field of ~ 100–200 V/cm results in a quenching of both lines. This quenching is due to a decrease of electron concentration near the heterointerface as a result of a photocurrent-induced heating of electrons in the GaAs layer. On the contrary, electrons near the heterointerface are effectively cooled, so the donors near the interface are not completely empty up to ~ 100 V/cm, which is in stark contrast with the case of bulk materials. The optical spin polarization of the donor-bound electrons near the heterointerface weakly depends on the electric field. Their polarization kinetics is determined by the spin dephasing in the hyperfine fields of the lattice nuclei. The long spin memory time (> 40 ns) can be associated with suppression of the Bir–Aronov–Pikus mechanism of spin relaxation for electrons.

## I. INTRODUCTION

More than 30 years ago, the attention of scientists was attracted by the mysterious H-band [1] of photoluminescence (PL) of AlGaAs/GaAs heterostructures. In the PL spectrum of GaAs it occupies an intermediate position between the X-band of the free exciton (1.515 eV) as well as exciton-impurity complexes near 1.514 eV and the shallow donor-acceptor-related band (D-A) near 1.493 eV. It was previously shown that the H-band is present in the PL spectra of both doped and undoped heterostructures [2], it has a long lifetime (~1–100 ns [1, 3, 4]) and disappears under slight temperature increase to above 10 K ($k_B T > 1$ meV, where $k_B$ is the Boltzmann constant and $T$ is the temperature) [1]. The H-band is absent in the PL spectra of bulk GaAs, which indicates its origin from the electronic states near the AlGaAs/GaAs interface.

It is known that the AlGaAs/GaAs heterojunction has an ideal interface and does not contain "intrinsic" Tamm states within the GaAs band gap [5] which can give rise to the H-band. Therefore, several models for the H-band were proposed. First, Ref. [6] considers a potential well for one type of carriers, which is formed in GaAs near the heterointerface due to band bending and results in either an electron or a hole accumulation channel (Figs. 1(a) and 1(b), correspondingly). For example, in the latter case, photoexcited holes are captured into the potential well at the heterointerface creating an excess positive charge. This positive charge attracts photoexcited electrons and results in formation of a potential 'dimple' ~ 1 meV deep. At low temperature photoelectrons are accumulated in this dimple. According to Ref. [6], hole and electron gases in the potential wells are degenerate, spatially separated, and confined along the growth direction. Second, in Refs. [7, 8] devoted to lightly doped heterostructures, the H-band was interpreted as recombination of spatially indirect excitons. In this model, charge carriers of one type, e.g. holes, are localized at the interface by the band bending and form spatially indirect excitons with charge carriers of the opposite sign (electrons). Finally, in Ref. [9] the H-band was associated with the recombination of donor-bound electrons with holes bound to interface acceptors ($D^0$–$A^0_{int}$).

Electric current flow through the photo-excited, moderately doped bulk GaAs strongly affects its edge PL [10, 11]. Quenching of the exciton-impurity PL was observed already for the very weak electric field of < 10 V/cm, which is much less than the ionization threshold of the exciton or neutral donor. The effect of electric current was associated with heating of electrons to the very high temperatures of up to 100 K for $E = 40$ V/cm [12] (the mobility of holes is much lower and thus they are not heated that much). Excitons and neutral donors are ionized by collisions with hot electrons or due to thermal ionization enhanced by the electric field [13]. Alternatively, heating of the electron gas decreases the probability of exciton formation from the electron-hole pairs and electron capture by a charged donors [13, 14]. As a result, the exciton-related channel of radiative recombination is strongly suppressed. The PL quenches due to the nonradiative recombination (previously ineffective as compared to the radiative one) being provided by surface states of GaAs. Ionization of donors in an electric field affects the spin polarization of the electron spin system [15]. The spin relaxation rate increases drastically due to the change in the spin relaxation mechanism from the hyperfine interaction with nuclei for the donor-bound electrons to the spin-orbit Dyakonov-Perel mechanism for the free electrons [16].

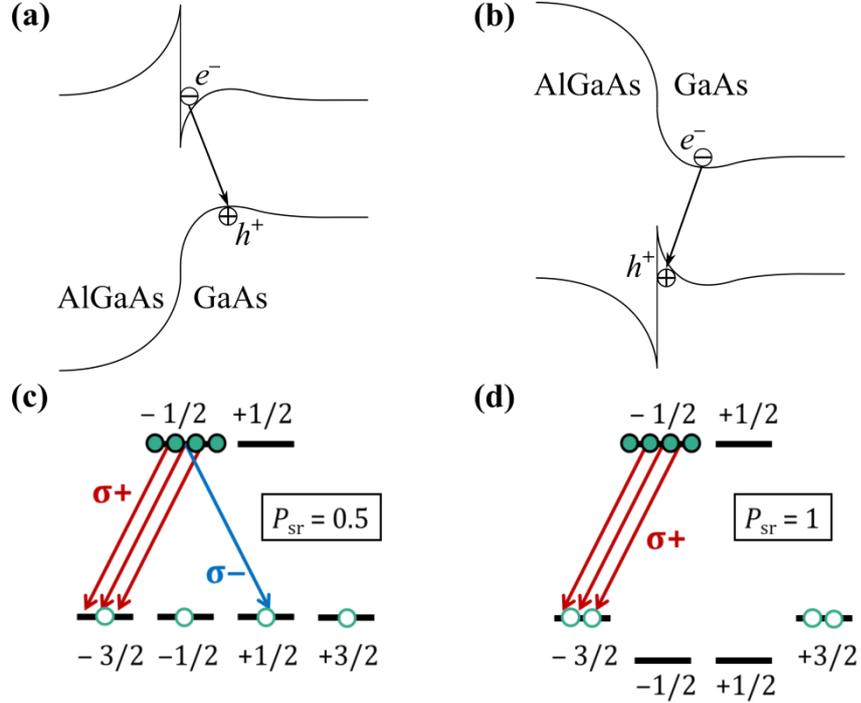

Figure 1. Schematic illustration of different models for the H-band: electrons (a) or holes (b) are localized in the potential well at the heterointerface and recombine with charge carriers of the opposite sign in the GaAs. Lower panels illustrate the corresponding selection rules for radiative recombination in case of degenerate (c) and split (d) heavy hole and light hole levels (electrons are fully polarized).

In case of an AlGaAs/GaAs heterostructure, the effect of the electron heating by an electric current is significantly different from that of bulk GaAs. Here, the role of the free surface and the related nonradiative recombination is minimized, which allows one to investigate the impact of the electric current heating on the GaAs-related PL in a wide range of electric fields. At low temperatures and low excitation levels, the H-band gives the main contribution to the GaAs PL spectrum. The influence of the electric current on the intensity and polarization of the H-band is still unexplored.

Here, we study optical orientation of the H-band in a nominally undoped AlGaAs/GaAs heterostructure and the effect of a weak, $E \ll 1$ keV/cm, lateral electric field, which is below the exciton ionization threshold in GaAs. Optical orientation measurements allow us to unambiguously distinguish among the above-mentioned mechanisms of the H-band formation and show that in the investigated structure a hole accumulation channel is realized (see Fig. 1(b)). Two lines with significantly different decay times constitute the H-band: a short-lived high-energy one and a long-lived low-energy one. The high-energy line originates from the recombination of holes from the accumulation channel and electrons freely moving along the structure plane in the dimple. The low-energy line is due to the recombination of the same holes with donor-bound electrons in the dimple. It is found that the lateral electric field changes the intensities of the two lines. With increasing electric field ($E > 100$ V/cm) the high-energy line, which has a shorter lifetime, disappears. A further increase in the electric field ($E > 200$ V/cm) decreases the intensity of the long-lived low-energy line. It is shown that the H-band quenching is due to a reduction of the charge carrier concentration near the heterointerface as a result of the current-induced heating of electrons in the GaAs layer. In stark contrast with bulk materials the donors near the interface are not completely empty up to $E \sim 100$ V/cm. The kinetics of the optical orientation of donor-bound electrons weakly depends on the electric field in this range and is determined by the spin dephasing in the hyperfine fields of the lattice nuclei. No change of spin relaxation mechanism is found for the H-band in the electric field. This indicates that, while electrons in the bulk GaAs layer are heated by the electric current, electrons near the heterointerface are not affected. The reasons for the effective energy relaxation near the interface are discussed.

## II. EXPERIMENTAL DETAILS

The AlGaAs/GaAs heterostructure under study is grown on a semi-insulating (100)-oriented GaAs substrate by molecular-beam epitaxy. It consists of ~ 1 µm $Al_{0.35}Ga_{0.65}As$, followed by a quasi-bulk (100 nm) GaAs layer covered with a 25 nm $Al_{0.35}Ga_{0.65}As$ cap layer (Fig. 2(a)). The concentration of shallow impurities is estimated to be $N_A - N_D < 10^{15}$ cm$^{-3}$, where $N_D$ and $N_A$ are the residual shallow donor and shallow acceptor concentrations, respectively. Two planar Au contact pads 1.5×1.5 mm are deposited on top of the structure. The distance $d$ between the contacts is 300 µm. The electric field strength $E$ is estimated as the ratio of the applied voltage $U$ to the distance $d$. The current-voltage characteristics of the structure is close to linear in the investigated range of applied voltages. At $E = 100$ V/cm the dark current is ~ 50 µA and the photocurrent of ~ 150 µA appears under continuous-wave (cw) excitation with the laser excitation at 1.69 eV with power density of ~ 10 mW/cm$^2$.

The PL in the cw regime is excited by a Ti:sapphire laser with photon energy of 1.69 eV, which is smaller than the $Al_{0.35}Ga_{0.65}As$ barrier bandgap energy (~2.0 eV). In order to eliminate the effect of dynamic nuclear polarization, the polarization of the laser beam is modulated between σ⁺ and σ⁻ by a photo-elastic modulator (PEM) operated at a frequency of 42 kHz. The excitation power density can be varied in the range from 1 mW/cm² to 1 W/cm². The PL signal is using a single grating monochromator equipped with a charge-coupled device (CCD) camera and an avalanche photodiode in the photon counting regime. The degree of the circular polarization of the PL [17] is obtained as follows: $\rho_{\sigma+}^c = \frac{I_+^+ - I_+^-}{I_+^+ + I_+^-}$, where $I_+^+$ and $I_+^-$ are the intensities of the σ⁺-polarized PL component under σ⁺ and σ⁻ excitation, respectively. PL measurements are carried out in longitudinal magnetic field $B_F$ parallel to the growth axis (Faraday geometry) or transverse magnetic field $B_V$ (Voigt geometry) up to 200 mT.

Time-resolved photoluminescence (TRPL) measurements are performed using a mode-locked Ti:sapphire laser (pulse duration of 150 fs, repetition frequency of 75.75 MHz). The central photon energy is 1.69 eV, and the spectral width of the laser pulses is about 15 meV. The average pump density is ~ 3 mW/cm². The laser beam is circularly polarized. The PL is dispersed with a 0.5 m focal length single monochromator and detected by a streak camera. The time resolution of the registration system is either 20 ps for measurements in the 2.5 ns time range, or 1 ns for the time range of 100 ns. In the latter case, a pulse picker is used to reduce the laser repetition rate so that the repetition period is about 1 μs. The TRPL measurements are conducted at zero magnetic field. The degree of circular polarization is obtained as follows: $\rho_c^{\sigma+} = \frac{I_+^+ - I_-^+}{I_+^+ + I_-^+}$, where $I_+^+$ and $I_-^+$ are the intensities of the σ⁺ and σ⁻ polarized PL components under σ⁺ excitation, respectively. The spectral shift of the PL spectra in longitudinal magnetic fields $B_F$ up to 9 T is measured using a CCD camera under linearly polarized excitation. All PL measurements (cw PL and TRPL) are performed at the temperature of $T = 2$ K.

### III. EXPERIMENTAL RESULTS
#### 1. Origin of H-band

The PL spectrum of the AlGaAs/GaAs heterostructure measured under cw excitation with the power density of 10 mW/cm² and at temperature $T = 2$ K is shown in the upper panel of Fig. 2(b). The spectrum consists of several bands: the exciton (X) band at 1.515 eV; the H-band, which is located 10 meV below the X-band; and the D–A band at 1.485–1.495 eV. The H-band consists of two lines, in agreement with Ref. [18]: a high-energy (HE) one and a low-energy (LE) one (Fig. 2(b), upper panel). There is also a low-energy shoulder at $\hbar\omega > 1.5025$ eV. The ratio of the intensities of the HE- and LE-lines strongly depends on the pump power and temperature.

In bulk unstrained GaAs-type semiconductors under optical orientation conditions, the selection rules for optical transitions are such that right-polarized (σ⁺) light excites three times more electrons in the conduction band with spin down (–1/2) than with spin up (+1/2) [16]. Thus, the spin polarization of the electrons at the moment of excitation is $P_e = \frac{n_{-1/2}-n_{+1/2}}{n_{-1/2}+n_{+1/2}} =$ 50% (where $n_{\pm 1/2}$ is concentration of photoelectrons with spin ±1/2). In case of non-resonant excitation, photoexcited holes quickly lose their polarization due to the strong spin-orbit interaction in the valence band. Radiative recombination of polarized electrons with non-polarized holes (Fig. 1(c)) is governed by the same selection rules as for absorption, and gives rise to the circular polarization of the PL. Thus, at the exciton transition (X) the maximum degree of circular polarization $\rho_{max} = P_e \cdot P_{sr}$ does not exceed 25%, as $P_{sr} = 0.5$ is given according to the selection rules (cf. Fig. 1(c)). A similar situation holds for the case, when photoelectrons are captured in the potential well at the heterointerface and recombine with photoexcited holes in GaAs (Fig. 1(a)).

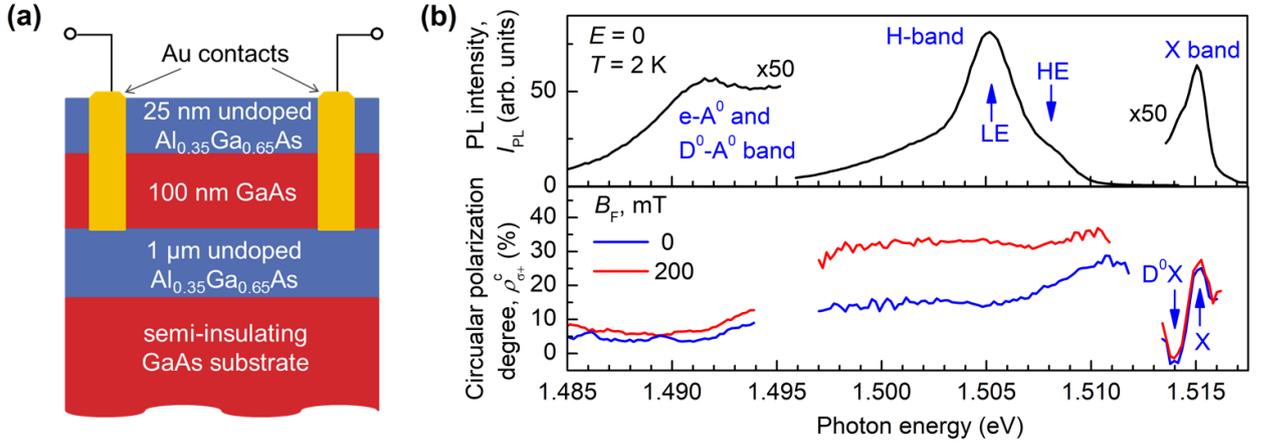

Figure 2. (a) Schematic representation of the AlGaAs/GaAs heterostructure. (b) Spectra of the intensity (upper panel) and degree of circular polarization (lower panel) of the PL in the longitudinal magnetic fields $B_F = 0$ and 200 mT ($E = 0$).

The lower panel in Fig. 2(b) shows the spectral dependences of the PL circular polarization degree $\rho^c_{\sigma+}$ in the optical orientation mode at zero field and in the longitudinal magnetic field of 200 mT. The circular polarization degree measured at the X-transition is indeed close to 25%. It is worth noting that the polarization degree drops to zero at the $D^0X$ transition since there the electron spins are antiparallel and the overall spin is zero. The polarization degree at the H-band is quite high – at the level of ~15%. A magnetic field in Faraday geometry ($B_F$) increases the optical orientation of electrons, suppressing their spin relaxation [16]. In $B_F = \pm 200$ mT, the degree of the circular polarization of the H-band increases to $\rho_{max}$ ~ 40%. It is obviously larger than the 25%-limit indicated above. This means that here the selection rules are different from the case of bulk GaAs or GaAs/AlGaAs heterointerface with an electron accumulation layer. Such a high degree of circular polarization can only be achieved if electrons recombine with *heavy* holes with a momentum projection of ±3/2, while the light hole states ±1/2 have a higher energy and are not populated at low temperatures. Indeed, the recombination of the photoexcited electrons and interface heavy holes is governed by selection rules with $P_{sr} = 1$, as illustrated in Fig. 1(d). It is known, that near the heterointerface the levels of heavy and light holes are split [19]. This means that in our structures holes, and not electrons, are localized at the heterointerface, either in the potential well formed by the band bending or at acceptors.

In Ref. [20] the composite structure of the H-band was identified by analyzing its spectral shift in a longitudinal magnetic field. The data below are consistent with the conclusions of that work. Figure 3(a) shows PL spectra of the investigated AlGaAs/GaAs heterostructure in longitudinal magnetic fields from 0 to 9 T. The position of the HE-line shifts linearly with magnetic field with a slope of 1.0 meV/T (Fig. 3(b)). The linear dependence indicates that the HE-line is due to recombination of electron-hole pairs which are not bound in exciton. The slope corresponds to the interband transitions between the zero Landau levels for electrons and holes: $\hbar\omega_{\mathrm{HE}}(B_F) = \hbar\omega_{\mathrm{HE}}(0) + \frac{\hbar e B_F}{2\mu}$, where $\hbar\omega_{\mathrm{HE}}(0)$ is the HE-line energy at zero magnetic field, $e$ is the electron charge, and $\mu = (1/m_\mathrm{e} + 1/m_\mathrm{h})^{-1}$ is the reduced effective mass. In GaAs the electron effective mass is $m_\mathrm{e} = 0.07 m_0$ (where $m_0$ is the free electron mass), then the coefficient $\frac{\hbar e}{2\mu} = 1.0$ meV/T corresponds to the heavy hole effective mass of $m_\mathrm{h} \approx 0.3 m_0$. This value is reasonable for the effective mass of the interface heavy hole in the plane of the heterointerface [20]. Thus, the HE-line corresponds to recombination of an electron and a hole, freely moving in the plane of the heterointerface.

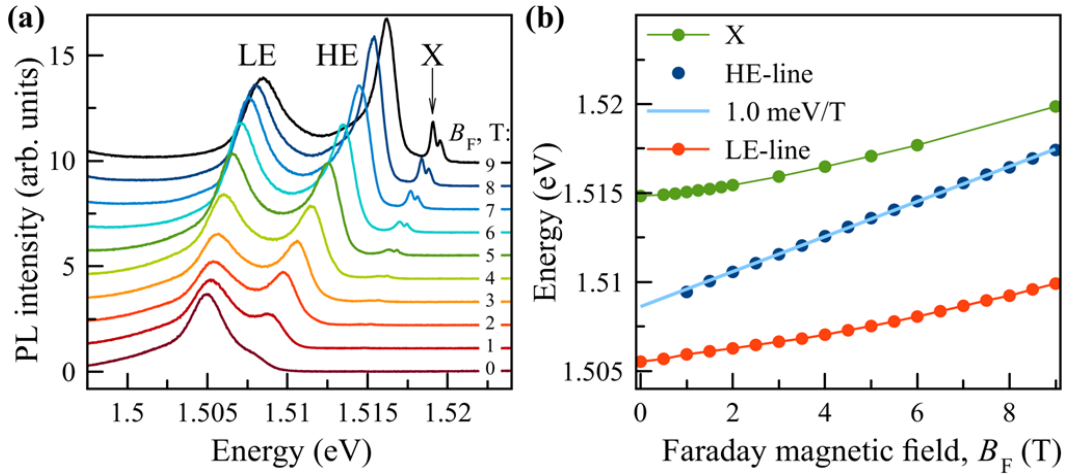

Figure 3. (a) The PL spectra of the H-band in longitudinal magnetic fields $B_F$ from 0 to 9 T for $\sigma^+$ detection. For clarity the spectra are shifted vertically. (b) Dependences of the spectral positions of the exciton peak (X), the HE- and the LE-lines on the longitudinal magnetic field $B_F$. The straight line is a linear fit for the HE-line position.

The magnetic field dependence of the energy position of the LE-line is nonlinear and similar to that of the bulk exciton (X) (Fig. 3(b)). However, most likely, the LE-line is not an exciton: since its binding energy is much smaller than that of the bulk exciton [7], its magnetic shift should be larger. We thus attribute the LE-line to the recombination of a donor-bound electron with a hole at the heterointerface, which is in accord with Ref. [18]. This interpretation is also supported by the time-resolved measurements of the spin polarization (see below), which was not studied previously. Its kinetics is characteristic of the dephasing of the electron in the hyperfine fields of the lattice nuclei [21].

Thus, from our results it follows that in the investigated heterostructure the H-band is formed according to the scheme shown in Fig. 1(b).

## 2. Effect of electric current on intensity and cw optical orientation of H-band

We consider now the influence of the in-plane electric field, which is much smaller than the ionization threshold of the GaAs exciton. Application of the electric field shifts the H-band spectrum to the lower-energy range (Fig. 4(a)). This indicates that the electric field results in an increase in the electron and hole separation, and thus should enhance the H-band decay time. Moreover, fields exceeding 100 V/cm are enough to completely eliminate the HE-line from the PL spectrum (see Fig. 4(a)), and in fields $E > 200$ V/cm the LE-line also vanishes (not shown).

The increase in the PL decay time, indeed, follows from the measurements of the Hanle effect. In transverse magnetic field $B_V$ (Voigt geometry), the stationary projection of the electron spin on the beam direction decreases due to the Larmor precession with the frequency $\omega_L = \mu_B |g| B_V/\hbar$, where $\mu_B$ is the Bohr magneton, $\hbar$ is the Planck constant, and the electron $g$-factor in GaAs is $g = -0.44$ [22]. This leads to PL depolarization, i.e. the Hanle effect [16]. The dependence of the circular polarization degree on the magnetic field has a Lorentzian form:

$$\rho_c(B_V) = \frac{\rho_0}{1+(B_V/B_{1/2})^2}, \quad \rho_0 = \rho_{max}\frac{T_s}{\tau} = \rho_{max}\frac{\tau_s}{\tau+\tau_s}. \qquad (1)$$

The maximum degree of circular polarization $\rho_{max}$ is determined by the selection rules and, as indicated above, is close to 40%. The half-width at half maximum (HWHM) of the depolarization curve is $B_{1/2} = \hbar/\mu_B g T_s$, where the electron spin lifetime $T_s$ is determined by its lifetime $\tau$ and spin relaxation time $\tau_s$:

$$1/T_s = 1/\tau + 1/\tau_s. \qquad (2)$$

The equations above are valid for free or weakly localized electrons. In case of strong electron localization (e.g., the distance between donors is much larger than the Bohr radius of an electron), there is no dynamic averaging of the Overhauser nuclear fields, and the situation is different. The dephasing of the electron spin in static random nuclear fields contributes to the width of the magnetic depolarization curve [21]. Generally speaking, in this case, the magnetic depolarization curve is no longer described by a Lorentzian. However the influence of these fields can be considered phenomenologically if the expression for the HWHM is represented as follows:

$$B_{1/2} = \hbar/\mu_B g \left(\frac{1}{T_s} + \Delta\omega_L\right). \tag{3}$$

The first term in the bracket in Eq. (3) is still determined by the spin lifetime $T_s$, but now the time $\tau_s$ is determined by interactions other than the hyperfine one. The second term describes the characteristic spread of Larmor frequencies $\hbar\Delta\omega_L = g\mu_B \Delta B_N + \Delta g \mu_B B_{1/2}$ due to the spread of random nuclear fields $\Delta B_N$ [21,23,24] and the spread $\Delta g$ in the electron $g$-factors. Here, however, the second term can be neglected since it does not affect the shape of the depolarization curve if $\Delta g \ll g$ (for our sample this condition is valid, see Appendix).

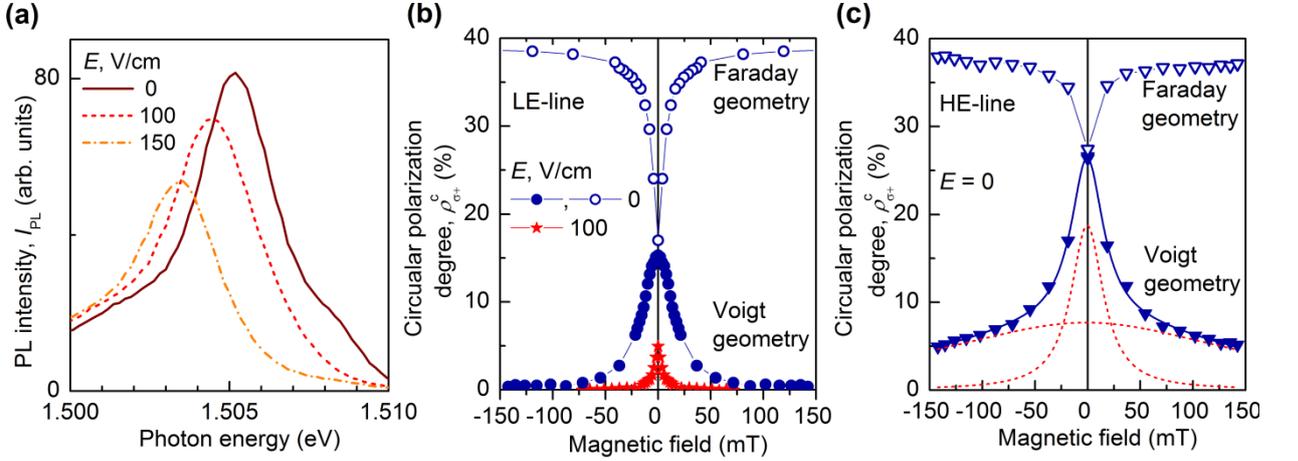

Figure 4. (a) H-band spectra at different electric fields. (b) Magnetic depolarization curves in the Voigt geometry at $E = 0$ and 100 V/cm and polarization recovery curve in the Faraday geometry ($E = 0$), measured at 1.505 eV (LE-line). (c) Magnetic depolarization curve in the Voigt geometry and polarization recovery curve in the Faraday geometry ($E = 0$), measured at 1.508 eV (HE-line). $T = 2$ K. Laser excitation at 1.69 eV photon energy with the power density of 10 mW/cm$^2$.

The PL depolarization curves measured at the LE-line (1.505 eV) in transverse magnetic field $B_V$ are shown in Fig. 4(b). At zero electric field (full blue circles) full depolarization is observed with $B_{1/2} = 15$ mT. An external in-plane electric field leads to narrowing of the magnetic depolarization curve: $B_{1/2} = 5$ mT at $E = 100$ V/cm. The electric field also reduces the spin polarization in zero magnetic field $\rho_0$ from 15% to 5%. It is worth noting that the PL circular polarization degree and depolarization curves measured at the low-energy shoulder ($\hbar\omega < 1.5025$ eV) are identical to those of the LE-line, which indicates their similar origin. However, the depolarization curve measured at the HE-line (1.508 eV) at zero electric field is significantly different (Fig. 4(c)). Its zero field spin polarization $\rho_0 = 26\%$, it has a non-Lorentzian shape and can be fitted by the sum of two Lorentzians with very different HWHMs: 18 and 175 mT. This can be explained taking into account that the LE-line dominates in the PL spectrum and overlaps with the HE-line (see the dashed lines in the upper panel in Fig. 2(b)). The depolarization curve measured at the HE-line contains also information on the spin polarization of the LE-line. So the narrower Lorentzian ($B_{1/2} = 18$ mT) comes from the LE-line, while the wider one ($B_{1/2} = 175$ mT) is truly related to the HE-line. Then, the difference in $B_{1/2}$ of the two depolarization curves indicates that the HE- and LE-lines originate from two ensembles of electrons (free and donor-bound, as indicated above) and that the spin exchange averaging between them is inefficient [25]. It is difficult to measure the HE-line depolarization curve at $E \neq 0$, as the line completely disappears from the cw spectra already at 100 V/cm.

Equations (1-3) show that the simultaneous decrease of the spin polarization in zero magnetic field and narrowing of the magnetic depolarization curve with increasing electric field for the LE-line are mainly due to the increase of the electron lifetime. The smallest HWHM of 5 mT obtained here corresponds well to the spread of the random nuclear fields of $\Delta B_N = 5.4$ mT evaluated in Refs. [21, 23]. TRPL data confirm this conclusion and show that the spin dephasing mechanism induced by random fields is robust with respect to electric fields of $E < 200$ V/cm (see below), in contrast to the case of bulk GaAs.

3.     **Effect of electric current on time-resolved polarized PL of H-band**

The TRPL technique is used for independent measurements of the intensity and polarization kinetics of the H-band. The slow decay of the PL intensity of the H-band (Fig. 5) in comparison with the free exciton lifetime in bulk GaAs (~ 100 ps) indicates that the electrons and holes are spatially separated. This is consistent with the scheme in Fig. 1(b) representing recombination of holes localized at the interface and attracted photoexcited electrons [1,2,6,9]. The PL intensity decay times of the HE- and LE-lines are quite different (Fig. 5(a)). This allows us to isolate their kinetics even in the absence of good spectral resolution. The PL intensity kinetics of the H-band is non-exponential possibly because of a change in the band bending with time during the recombination of photocarriers after arrival of a laser pulse. The band bending determines the electron and hole wave function overlap and hence the probability of their recombination. Although the decay of both lines is not exponential, we take below the time at which PL intensity decreases by a factor of $e = 2.718...$ for estimating the PL decay time, The high-energy line HE decays with a characteristic decay time $\tau_{HE} \approx 200$ ps (Fig. 5(b)). Its intensity amplitude $I_{HE}(0)$ decreases in an electric field, while the decay time $\tau_{HE}$ does not change. The intensity kinetics of the LE-line is much slower with the characteristic decay time $\tau_{LE} \sim 5$ ns (Fig. 5(c)).

Application of the lateral electric field results in a decrease of the LE-line intensity $I_{LE}(0)$ as well as in an increase of its decay time (up to $\tau_{LE} \sim 10$ ns at $E = 100$ V/cm).

The kinetics of the circular polarization degree of the LE-line demonstrates a decay with a characteristic time of ~ 5 ns followed by saturation at the level of $\rho_c^{\sigma+} \approx 15$–$20\%$ for at least 40 ns (Fig. 5(d)). Such non-exponential kinetics cannot be described by the spin relaxation time $\tau_s$ and is typical for spin dephasing of the donor-localized electrons in random hyperfine fields of unpolarized nuclei [21]. In this case the initial decay is characterized by the spin dephasing time $T_2^* = \hbar/g\mu_B\Delta B_N$. For a shallow donor $\Delta B_N$ is around 5 mT [22] resulting in $T_2^* \approx 5$ ns in full agreement with the experimental results. So it can be assumed that the electrons contributing to the LE-line are localized on shallow donors near the heterointerface.

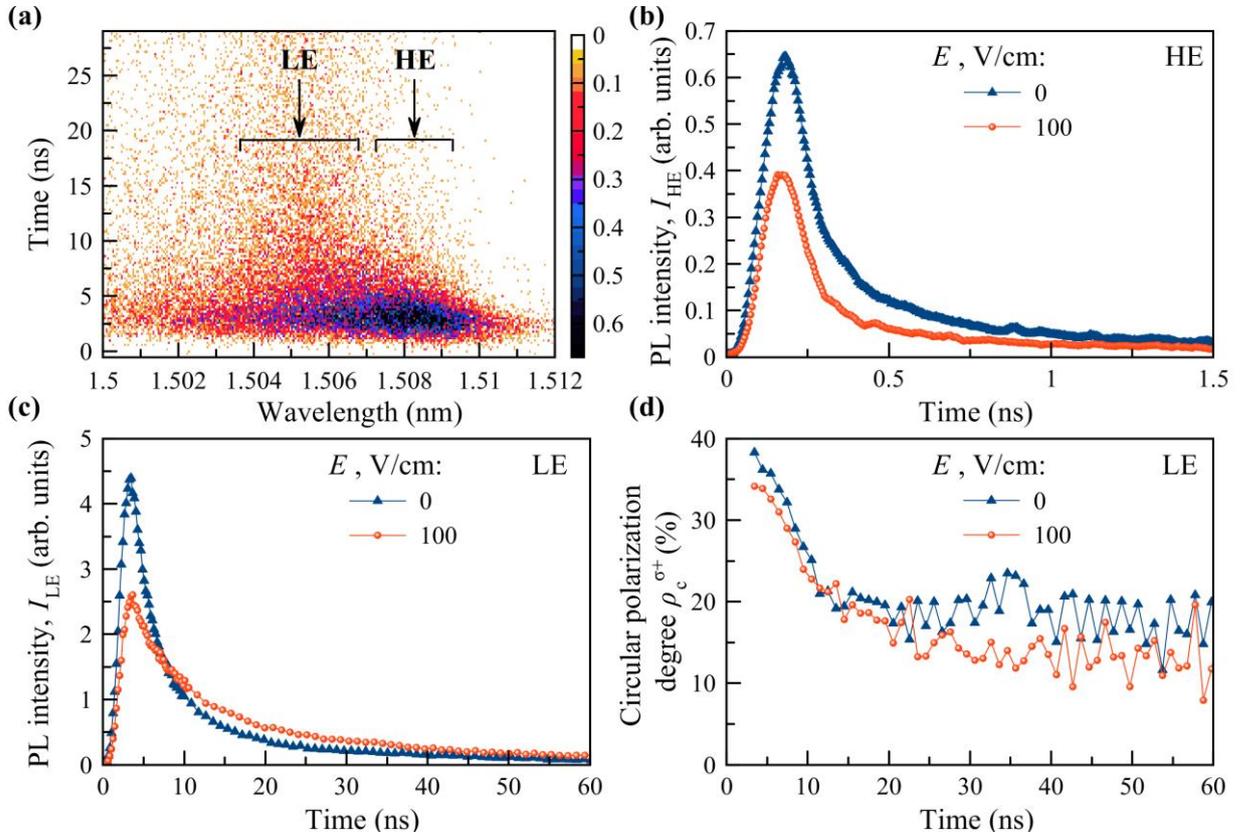

Figure 5. PL intensity and polarization kinetics. (a) Spectral dependence of PL transients within the H-band ($E = 0$). (b) Intensity kinetics of the HE-line at various electric fields; the maximum intensity is normalized to unity. (c) The same as (b), but for the LE-line. (d) The degree of the circular polarization $\rho_c^{\sigma+}(t)$ of the LE-line. $T = 2$ K.

The long spin memory time (plateau for > 40 ns, see Fig. 5(d)) can be associated with the suppression of the Bir–Aronov–Pikus mechanism [16] of spin relaxation of electrons: the exchange interaction of the electrons with the interface holes is suppressed due to their spatial separation. It is interesting that the application of an electric field up to 100 V/cm does not affect the polarization kinetics. Hence the mechanism of spin relaxation is robust to the electric field, in contrast with bulk GaAs [15].

It is worth noting that in Ref. [9], in which the H-band is associated with the $D^0$–$A^0_{int}$ recombination, the PL decay time scales exponentially with the PL photon energy within the H-band. This is due to the spatial distribution of the $D^0$–$A^0_{int}$ pairs. A smaller $D^0$–$A^0_{int}$ distance corresponds to a higher probability of radiative recombination, as well as a stronger Coulomb interaction, and thus a smaller PL photon energy. This is in striking contrast with our results, as within the LE-line and its low-energy shoulder there is no spectral dependence of the decay time. Therefore, in the structure investigated here the LE-line cannot be related to $D^0$–$A^0_{int}$ recombination.

## IV. DISCUSSION

Our experiments show that the H-band in the studied AlGaAs/GaAs heterostructure arises from recombination of heavy holes in the quantum well at the heterointerface with electrons in the GaAs layer, as sketched in Fig. 1(b). The quantum well for holes is formed by the band bending as suggested in Ref. [6]. The depth of the quantum well should be ~15 meV so that the energy of the recombining hole and electron corresponds to the spectral position of the H-band. The necessary band bending may initially come from the depletion of shallow or deep acceptors in the AlGaAs barrier, which provides holes to form an accumulation layer at the heterointerface. The presence of the accumulation layer in equilibrium is in agreement with the nonzero dark current, which is ~ 50 μA at $E = 100$ V/cm (while the photocurrent under cw laser excitation at 1.69 eV with the power density of ~ 10 mW/cm$^2$ is ~ 150 μA).

Application of the electric field of ~ 100 V/cm in the plane of the structure results in the following experimental facts in the cw mode: (1) decrease in the integrated H-band intensity $I_{\text{H-band}} = I_{\text{HE}} + I_{\text{LE}}$; (2) decrease of the ratio of the HE- and LE-line intensities $I_{\text{HE}}/I_{\text{LE}}$; (3) low-energy shift of the H-band; (4) decrease in the circular polarization degree $\rho_0$ at the H-band and the depolarization curve HWHM $B_{1/2}$ in the optical orientation regime; (5) the HE-line corresponds to the recombination of electrons that move freely in the dimple in the plane of the heterointerface, while the LE-line is associated with the recombination of electrons bound to donors in the vicinity of the dimple. Time-resolved spectroscopy shows that the application of electric field also leads to (6) decrease in the intensity amplitudes of both lines, $I_{\text{HE}}(0)$ and $I_{\text{LE}}(0)$; (7) increase in the LE-line decay time $\tau_{\text{LE}}$, while (8) the decay time of the HE-line $\tau_{\text{HE}}$ does not change; (9) the non-exponential kinetics of the optical orientation $\rho_c^{\sigma+}(t)$ at the LE-line with its saturation for at least 40 ns is a characteristic of dephasing of electron spins in the hyperfine nuclear fields and remains unchanged in electric field.

All these results can be understood within the following simple model involving a minimal set of parameters. Figure 6 shows an AlGaAs/GaAs heterointerface with the band bending which leads to formation of a quantum well for holes on the GaAs side of the heterojunction. The H-band is excited non-resonantly, by excitation of electrons and holes in the quasi-bulk 100-nm GaAs layer. The photoexcited holes fall into the quantum well at the interface. Their excess charge creates a shallow Coulomb potential well – a 'dimple' – for electrons near the interface. The photoexcited electrons attracted to the interface by the excess positive charge of the captured holes, lose energy and fall into the dimple. The electrons in the dimple are confined only along the growth direction of the heterostructure, while they can move freely along the plane of the heterointerface. Radiative recombination of the electrons from the dimple with the heavy holes from the interface quantum well gives rise to the HE-line. A fraction of the electrons from the dimple is captured on shallow donors, and recombines with the same interface heavy holes. leading to the LE-line.

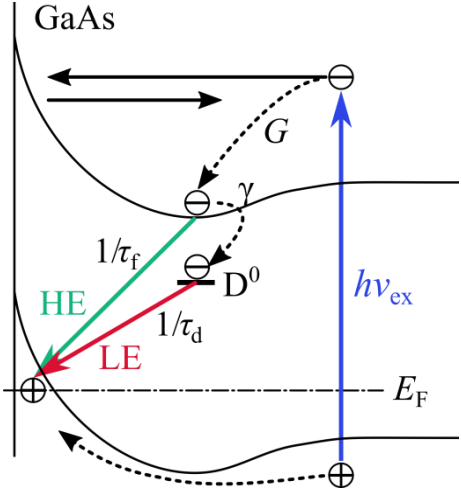

Figure 6. Sketch of the HE- and LE-lines formation in the studied heterostructure upon interband excitation.

Accordingly, one can write the following rate equations for the electrons in the dimple:

$$\frac{dn_f}{dt} = G - \gamma n_f - \frac{n_f}{\tau_f}; \quad \frac{dn_d}{dt} = \gamma n_f - \frac{n_d}{\tau_d}. \tag{4}$$

Here $G$ is the flux of electrons into the dimple, $n_f$ is the concentration of free electrons in the dimple, and $n_d$ – the concentration of electrons captured on donors. The radiative lifetimes of the free and donor-bound electrons in the dimple are given by $\tau_f$ and $\tau_d$, respectively, and $\gamma$ is the capture rate of electrons from the dimple onto empty donors. The PL intensities of the HE- and LE-lines are given by:

$$I_{\text{HE}} = \frac{n_f}{\tau_f}; \quad I_{\text{LE}} = \frac{n_d}{\tau_d}. \tag{5}$$

Let us consider the stationary and non-stationary solutions of Eqs. (4) for the cw and time-resolved modes, respectively. In the cw mode, the solutions of Eqs. (4) and (5) are the following:

$$n_f = G\frac{\tau_f}{1+\gamma\tau_f}; \quad n_d = G\tau_d\frac{\gamma\tau_f}{1+\gamma\tau_f}; \quad I_{\text{HE}} = G\frac{1}{1+\gamma\tau_f}; \quad I_{\text{LE}} = G\frac{\gamma\tau_f}{1+\gamma\tau_f}. \tag{6}$$

Thus, the overall PL intensity of the H-band is determined only by the electron flux into the dimple $I_{\text{H-band}} = I_{\text{HE}} + I_{\text{LE}} = G$ and does not depend on the temporal parameters $\tau_f$, $\tau_d$, and $\gamma$. It follows from the experimental data that with an increase in electric field the intensity of the H-band, and therefore the electron flux $G$ into the dimple, decrease. The relative intensity of the HE-line is $\frac{I_{\text{HE}}}{I_{\text{LE}}} = \frac{1}{\gamma\tau_f} \lesssim 0.2$ (see the upper panel in Fig. 2(b)) and also decreases with the increasing electric field. This means, that $\gamma\tau_f \gg 1$ and increases with electric field. Below, it will be shown that with increasing electric field $\gamma$ does not change and hence it is the radiative lifetime $\tau_f$ that should increase.

In the time domain, after the excitation pulse arrival ($t > 0$), Eq. (4) can be solved for $G = 0$ using the following initial conditions: $n_f(0) = G\tau_p \equiv g; \; n_d(0) = 0$, where $\tau_p$ is the excitation pulse duration. The decay time of the HE-line $\tau_{\text{HE}}$ is given by $(\gamma + \frac{1}{\tau_f})^{-1}$, and the rise time of the LE-line is $1/\gamma$. Both processes occur as fast as several hundreds of picoseconds, as can be seen from Figs. 5(b) and (c). This time is much shorter than the radiative lifetime $\tau_f$ because $\gamma\tau_f \gg 1$. In this case, the solutions of Eq. (4) can be written as:

$$I_{\text{HE}}(t) = \frac{g}{\tau_f}e^{-\gamma t}; \quad I_{\text{LE}}(t) = g\gamma\frac{e^{-\gamma t}-e^{-t/\tau_d}}{1-\gamma\tau_d}. \tag{7}$$

Hence, the decay of the HE-line is mostly governed by the nonradiative capture of free electrons by charged donors: $\tau_{\text{HE}} = 1/\gamma$. The long-term decay of the LE-line (at $t \gg \gamma^{-1}$) is determined by the radiative lifetime $\tau_{\text{LE}} = \tau_d$:

$$I_{LE} \approx \frac{g}{\tau_d}e^{-t/\tau_d} \tag{8}$$

The comparison with the experimental results allows us to extract the electric field dependence of the model parameters and give a physical interpretation of the processes under consideration. As it was shown above, the lateral electric field results in a decrease of the electron flux $G$ into the dimple. Thus means therefore that application of the electric field leads to a decrease in the concentration of electrons trapped at the heterointerface. The TRPL data show that the electric field does not affect the decay time $\tau_{\text{HE}}$ of the HE-line, but results in an increase of the decay time of the LE-line (see Figs. 5 (b) and (c)). So the capture $\gamma$ does not depend on the electric field. At the same time, the application of the electric field of ~ 100 V/cm results in an enhancement of the radiative lifetimes $\tau_f$ and $\tau_d$ of both lines.

The decrease of the electron flux $G$ and simultaneous increase of the radiative lifetimes is consistent with the low-energy Stark shift of the H-band in the lateral electric field (see Fig. 4(a)). It is known that photogenerated electron-hole pairs are separated at the heterointerface and screen the built-in electric field of the heterojunction, thus flattening the bands. The decrease in the concentration of electrons trapped in the heterointerface dimple should lead to a reduction of the screening effect. So application of the *in-plane* electric field indirectly results in an enhancement of the band bending and thus an increase of the spatial separation of electrons and holes at the heterointerface.

The data on the Hanle effect are also in qualitative agreement with this picture. Indeed, the application of the electric field decreases the circular polarization degree $\rho_0$ of the LE-line from 15 % to 5% and simultaneously decreases the HWHM of the magnetic depolarization curve from 15 mT to 5 mT. Both facts are consistent with the increase of the LE-line lifetime (see Eqs. (1-3)).

Earlier, the PL quenching in weak electric fields of ~10 V/cm (small enough as compared to the ionization threshold for excitons and donors) was well studied in lightly doped bulk GaAs. It was explained in terms of a strong heating of the electron gas by the electric current, which leads to a decrease in the probability of capture of the heated electrons by the Coulomb centers and/or an increase in the probability of ionization of the shallow impurities [14, 13]. Therefore, it is reasonable to assume a similar heating by the electric current of electrons in the quasi-bulk GaAs layer of the AlGaAs/GaAs heterostructure. These electrons are then captured at the heterointerface and give rise to the H-band. The current-induced heating leads to a decrease of the flux $G$ of the electrons captured in the dimple. This results in the Stark shift of the H-band and the enhancement of the radiative lifetimes $\tau_f$ and $\tau_d$ of both emission lines, in agreement with our experimental results.

However, there is a significant difference from the bulk situation: the capture probability by a charged donor in the dimple should decrease if the electrons near the interface are heated as in bulk GaAs [13]. In the heterostructure investigated here this is not the case. As it was shown above, the capture rate $\gamma$ does not depend on the in-plane electric field. Moreover, the spin dynamics of the LE-line, which corresponds to the dephasing of the spin of a donor-bound electron near the interface, is preserved for electric fields as large as 100 V/cm. At the same time, in pure bulk GaAs, the depletion of donors takes place already in fields of around 10 V/cm and leads to a strong acceleration of the electron spin relaxation [15] due to the change in the spin relaxation mechanism from the hyperfine interaction with nuclei for the donor-bound electrons to the spin-orbit Dyakonov-Perel mechanism for the free electrons.

All these results force us to assume that, while electrons in the quasi-bulk GaAs layer are heated by the electric current, those in the dimple are not. In other words, the heated electrons in the bulk that are captured in the dimple experience effective energy relaxation. The relaxation can be due to the emission of phonons with different wave vectors (near the interface the conservation law of momentum along the growth axis of the structure is relaxed). Another reason can be the cooling of electrons by inelastic scattering on the holes of the accumulation layer via a so-called shake-up process [26]. Finally, the mobility of the electrons in the dimple may be decreased due to scattering on the fluctuations of the dimple profile.

## V.    CONCLUSION

We have studied the effect of a weak in-plane electric current on the PL intensity and optical orientation of the H-band in a lightly doped AlGaAs/GaAs heterostructure. Optical orientation measurements show that the case of a hole accumulation channel is realized in this structure. The H-band is formed due to the attraction of electrons and holes from the quasi-bulk GaAs layer to the heterointerface. The photoexcited holes fall into the quantum well at the interface and their excess charge creates a shallow Coulomb potential well – a 'dimple' – for electrons near the interface. Two lines with significantly different decay times constitute the H-band: the short-lived high-energy one and the long-lived low-energy one. The high-energy line originates from the recombination of electrons moving freely along the interface plane. The low-energy line is due to the recombination of donor-bound electrons.

It has been found that the lateral electric field modifies the intensities of the two lines. With an increase in the electric field (>100 V/cm) the high-energy line disappears. This quenching is due to the reduction of the charge carrier concentration near the heterointerface as the result of current-induced heating of the electrons in the GaAs layer. It has been also shown that the capture rate of electrons from the dimple onto empty donors does not depend on the electric field.

The spin dynamics measured at the low-energy line is typical for electrons localized on shallow donors. In stark contrast with bulk materials, the donors near the heterointerface are not completely empty in electric fields of up to 100 V/cm. The kinetics of their optical orientation weakly depends on the electric field in this range and is determined by the spin dephasing in the hyperfine fields of the lattice nuclei. No change of the spin relaxation mechanism by the electric field is found at the low-energy line. After spin dephasing during the first ~10 ns the spin polarization of the electrons does not change for at least 40 ns. The fact that both the spin polarization kinetics and the capture rate of the electrons from the dimple onto the donors are independent on the electric field suggests that, while electrons in the bulk GaAs layer are heated by the electric current, the electrons in the dimple are not.

**Acknowledgements**

We acknowledge support by the Russian Science Foundation, grant #18-12-00352 (O.S.K., V.L.K., I.V.K., V.F.S., Yu.G.K.). E.A.Zh., I.A.A., A.D.W., A.L., D.R.Y. and M.B. thank the Deutsche Forschungsgemeinschaft and the Russian Foundation of Basic Research for support in the frame of the ICRC TRR 160 (projects A1 and B4). A.D.W. and A.L. also thank the grants DFH/UFA CDFA05-06, DFG project 383065199, and BMBF Q.Link.X 16KIS0867.

**Appendix**

**Effect of electric current on pump-probe Kerr rotation.**

The pump-probe time-resolved degenerate Kerr rotation (TRKR) technique [27] is used for studying the coherent spin dynamics of the electrons in magnetic field. The light beam from a mode-locked Ti:sapphire laser (pulse duration of 1.5 ps and repetition period of 13.2 ns) is divided into the pump and probe beams. The polarization of the pump is modulated at 50 kHz between σ⁺ and σ⁻ by a photoelastic modulator PEM. The pump-induced spin coherence is detected in reflection geometry through the Kerr rotation of the linear polarization of the probe beam. The spot diameter of the pump beam on the sample is 350 μm, while the spot diameter of the probe beam is slightly smaller. The probe and pump beam powers are in the range of 0.5–1.0 and 1–10 mW, respectively. The photon energy of the pump and probe pulses are identical and tuned in resonance with the free exciton states at 1.515 eV. The TRKR signals are measured in transverse (perpendicular to the photon wave vector and parallel to the plane of the sample) magnetic field $B_\text{V}$ up to 3 T. The time resolution of the registration system is 1.5 ps. The angle of the Kerr rotation of the polarization plane of the probe beam is proportional to the $z$-component of the spin polarization. TRKR allows us to measure the spin dephasing time ($T_2^*$) of the electrons, the frequency of Larmor precession ($\omega_\text{L}$) of the electron spins (estimate the $g$-factors), and also to estimate the spread of the $g$-factor ($\Delta g$). TRKR signals are recorded at the temperature of 1.7 K.

The TRKR signals in the transverse magnetic field $B_\text{V} = 200$ mT with the in-plane electric field $E$ in the range from 0 to 170 V/cm are shown in Fig. 7(a). The signal can be approximated by a damped oscillatory function arising from precession of spins in the transverse magnetic field:

$$\theta_\text{K}(t) = \Theta \cos(\omega_\text{L} t) \exp(-t/T_2^*), \tag{4}$$

where $\theta_K(t)$ is the TRKR signal, $\Theta$ is the oscillation amplitude, $t$ is the delay time between the pump and probe pulses, and $T_2^*$ is the spin dephasing time (decay time of the spin oscillations). The $g$-factor determined from the oscillation frequencies of the Larmor precession is $|g| = 0.435$, which indicates that in this case the electron spin dynamics is measured. At $E = 0$ and $B_V = 200$ mT, the spin dephasing time $T_2^*$ is ~ 1.5 ns. An increase in the in-plane electric field leads to an increase in $T_2^*$, e.g., at $E = 170$ V/cm time $T_2^*$ reaches to ~ 4.5 ns.

In order to estimate the spread of $g$-factors ($\Delta g$) the dependence of $1/T_2^*$ on the transverse magnetic field $B_V$ is measured at zero electric field at the temperature of 1.7 K (Fig. 7(b)). The measured dependence to a good approximation can be described by the following expression [28]:

$$1/T_2^* = 1/T_s + \Delta\omega_L = 1/T_s + \frac{1}{\hbar}g\mu_B\Delta B_N + \frac{1}{\hbar}\Delta g\mu_B B_V. \quad (5)$$

It allows us to obtain the $g$-factor spread $\Delta g \approx 0.025$ for $T = 1.7$ K. So the spread of the $g$-factors is small compared to its absolute value, $\Delta g \ll g$.

Therefore, the TRKR data do not contradict the PL data described in the main text, although the TRKR technique measures the spin dynamics of electrons under pumping into the exciton resonance and thus reflects the spin dynamics at the H-band only indirectly.

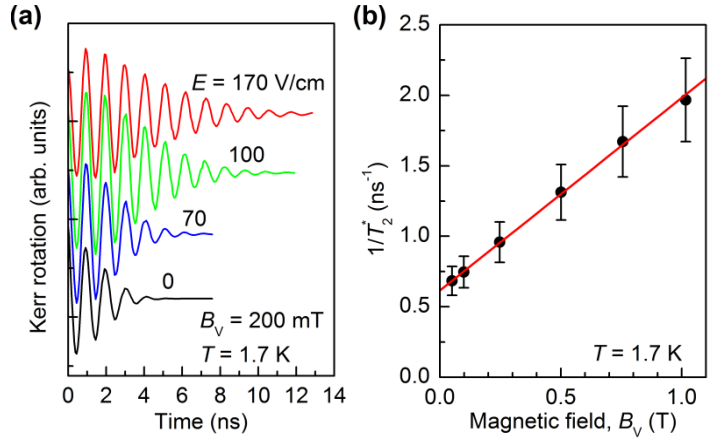

Figure 7. (a) TRKR signals in the transverse magnetic field $B_V = 200$ mT at different electric fields. (b) Magnetic dependence of $1/T_2^*$ at $E = 0$. The power density of the pump and probe beams are 3 and 1 mW, respectively. $T = 1.7$ K.